\documentclass{elsart}

\usepackage{graphicx}

\newcommand{\tr}{\mbox{tr}}

\newcommand{\bra}[1]{\langle#1|}
\newcommand{\ket}[1]{|#1\rangle}

\newcommand{\ketbrad}[1]{|#1\rangle\!\langle #1|}
\newcommand{\eqref}[1]{(\ref{#1})}

\begin{document}

\begin{frontmatter}

\title{Quantum entanglement, indistinguishability, and the absent-minded driver's problem}
\author{Ad\'{a}n Cabello}
\ead{adan@us.es}
\address{Departamento de F\'{\i}sica Aplicada
II, Universidad de Sevilla, E-41012 Sevilla, Spain}
\author{John Calsamiglia}
\ead{John.Calsamiglia@uibk.ac.at}
\address{Institut f\"{u}r Theoretische Physik,
Universit\"{a}t Innsbruck, A-6020 Innsbruck, Austria}


\begin{abstract}


The absent-minded driver's problem illustrates that probabilistic
strategies can give higher pay-offs than deterministic ones. We
show that there are strategies using quantum entangled states that
give even higher pay-offs, both for the original problem and for
the generalized version with an arbitrary number of intersections
and any possible set of pay-offs.
\end{abstract}


\end{frontmatter}



\section{Introduction}


\subsection{The absent-minded driver's problem}

\begin{figure}
\centerline{\includegraphics[width=4.2cm]{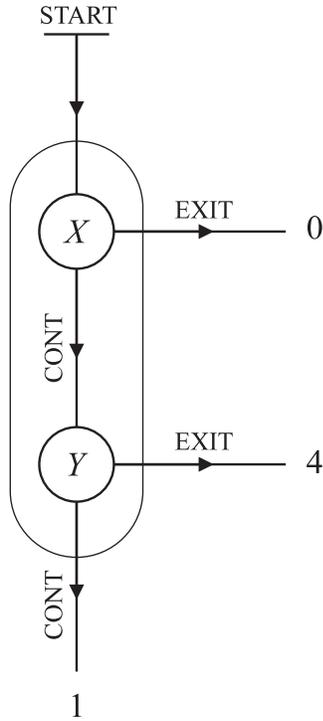}}
\caption{\label{Roadhome} The absent-minded driver's problem.}
\end{figure}

The so-called {\em paradox of the absent-minded driver} was
introduced by Piccione and Rubinstein in~\cite{PR97a} and further
discussed in~\cite{AHP97,PR97b} and references therein: an
individual is sitting late at night in a bar planning his midnight
trip home. The trip starts at the bar, the {\sc start} in
Fig.~\ref{Roadhome}. There is a highway with two consecutive exits
(or intersections), $X$ and $Y$, and he has to take the second,
$Y$, to get home (pay-off~4). If he takes the first one, he
arrives at a bad neighborhood (pay-off~0), and if he fails to take
either, he has to stay in a motel at the end of the highway
(pay-off~1). He cannot go back. There are two essential
assumptions:

\begin{itemize}
\item[(I)] {\em Indistinguishability}: The intersections $X$ and
$Y$ are indistinguishable by any experiment performed at one
intersection. When the driver is at one intersection, no
experiment can give him information about which intersection he is
at.

\item[(II)] {\em Absent-mindedness}: The driver is absent-minded
and is aware of this fact. Absent-mindedness only affects his
memories about whether he has already gone through one of the
intersections; at $X$ he knows he might be at $Y$ but has
forgotten passing $X$, and at $Y$ he cannot remember passing $X$.
Apart from this, the driver is perfectly able to remember anything
else.
\end{itemize}

Some remarks about these assumptions follow:

\begin{itemize}
\item[(i)] These assumptions are not independent: the
indistinguishability of the intersections is only relevant for an
absent-minded driver, and absent-mindedness is only relevant when
the intersections are indistinguishable (otherwise, the driver
could obtain information for decision-making in spite of his
absent-mindedness).

\item[(ii)] Since the driver's absent-mindedness is limited to his
memories about the intersections, but does not prevent him from
possessing information about the rest of the universe, then it is
forbidden any experiment at one intersection whose result,
together with any information about the rest of the universe,
allows the driver to obtain information about which intersection
he is at.

\item[(iii)] Implicit in the rules is the fact that the driver cannot
transmit information from one intersection to the rest of the
universe (and, in particular, to the other intersection), because
this could be used to distinguish the intersections.
\end{itemize}



The above scenario allows Piccione and Rubinstein to exhibit a
conflict between two ways of reasoning at an intersection:

\begin{quotation}
``Planning his trip at the bar, the decision maker must conclude
that it is impossible for him to get home and that he should not
exit when reaching an intersection. Thus, his optimal plan will
lead him to spend the night at the motel and yield a payoff of~1.
Now suppose that he reaches an intersection. If he had decided to
exit, he would have concluded that he is at the first
intersection. Having chosen the strategy to continue, he concludes
that he is at the first intersection with probability~1/2. Then,
reviewing his plan, he finds that it is optimal for him to leave
the highway since it yields an expected payoff of~2. Despite no
new information and no change in his preferences, the decision
maker would like to change his initial plan once he reaches an
intersection!~\cite{PR97a}''
\end{quotation}

Piccione and Rubinstein make use of this apparent paradox to
illustrate the advantages of probabilistic (or {\em
random}~\cite{PR97a}, or {\em mixed}) strategies versus
deterministic (or {\em pure}~\cite{PR97a}) strategies. At the
intersections, the driver can either {\sc continue} along or {\sc
exit} the highway. Accordingly, there are are two possible
deterministic strategies: either to always {\sc continue}
(pay-off~1) or to always {\sc exit} (pay-off~0). Alternatively, at
the intersections, the driver can toss a (suitable weighted) coin
with a probability $p$ for heads (which means {\sc continue}) and
a probability $1-p$ for tails (which means {\sc exit}). The
expected pay-off of this probabilistic strategy
is~$4\,p\,(1-p)+p^2$. Therefore, if $p>1/3$, this strategy gives a
higher pay-off than the best deterministic strategy. The optimal
probabilistic strategy consists of choosing $p=2/3$
(pay-off~$4/3$).


\subsection{Quantum strategies}


Game theory has recently found a new direction based on the
possibility of the resources of quantum mechanics becoming
available to the players~\cite{Klarreich01}. This introduces new
possibilities which lead to advantages over their classical
counterparts (however, see~\cite{VP02}). So far, the advantages of
the so-called quantum strategies have been limited to situations
involving players with perfect
recall~\cite{Meyer99,EWL99,BH01,KJB01,DLXSWZH02,LJ02,GM03}. As we
shall illustrate, decision problems involving players with
imperfect recall are a promising arena for the application of
quantum resources. In particular, we will see that quantum
mechanics allows the absent-minded driver to make use of
correlations without compromising the assumptions (or rules) of
the problem and thereby obtain higher pay-offs.


\section{Physical realization of probabilistic strategies}


Let us start by re-examining from the perspective of physics what
Piccione and Rubinstein consider the optimal solution to the
absent-minded driver's problem: a probabilistic strategy. Can we
implement a probabilistic strategy in a real experiment without
compromising the assumptions of the problem at a fundamental
level? In particular, how can we accomplish the driver's action of
``tossing a coin'' in the intersection satisfying the requirement
of even-in-principle indistinguishability of the intersections?
Strictly speaking, the driver is not allowed to carry a coin since
he could use it to bypass his absentmindedness and keep track of
the intersections he passes. However, the driver may instead place
a coin in each of the intersections before the trip starts.

The first problem is which physical system can be used as a coin.
Strictly speaking, real coins are highly complex systems and
therefore, in principle, it is not difficult to distinguish
between two of them. This problem can be eluded by replacing real
coins by identical physical systems (i.e., those with the same
composition of identical elementary particles and prepared in the
same physical state).

The next problem is the mechanism for ``tossing'' the coin. By
``tossing a coin'' we mean that the driver can use his coin to
perform a two-output experiment with probability $p$ for one of
the outputs. However, this is a very difficult task from the
perspective of classical physics, where the results of experiments
are essentially predefined in the state of the coin and the device
used to toss it. In this scenario both are under the driver's
control, and therefore the result of the experiment can in
principle be fixed by the driver beforehand. In quantum mechanics,
however, it can be proven, under some general assumptions, that
the results of certain experiments are not
predefined~\cite{Bell64,GHZ89}, but are created~\cite{Wheeler78}
when the experiment is performed. Therefore, at least from the
perspective of quantum mechanics, there is a method for tossing a
coin without compromising the assumptions of the absent-minded
driver's problem at a fundamental level. The driver can place a
``quantum coin''~\cite{Meyer99} or qubit (i.e., a two-level
quantum system), say a spin-$1\over 2$ particle, in each of the
intersections ($X$ and $Y$). Let us denote the quantum state of
the two quantum coins as $\rho_{XY}$.

The indistinguishability assumption implies that the quantum state
of each quantum coin is represented by the same reduced density
matrix, $\tr_X \rho_{XY}=\tr_Y \rho_{XY}$.
Moreover, states with unequal reduced density matrices hidden in
classical mixtures (i.e., states of the form $\rho_{XY}=\sum_i p_i
\ketbrad{\psi^i_{XY}}$, where, in some instances, $i=a$, the
quantum coins are in different states,
$\tr_X\psi^a_{XY}\neq\tr_Y\psi^a_{XY}$) are forbidden, since there
is no way to ensure that the driver does not have a priori
knowledge on the particular instance that will appear in each
intersection in a particular trip. This knowledge would allow the
driver to learn, with a given probability, which intersection he
is at. Hence this does not fulfill the assumption of
indistinguishability.

A probabilistic strategy can be implemented right away by a suitable
measurement on a suitable quantum state. For instance, by
preparing the qubits in the pure state
\begin{equation}
\ket{\Phi}=\ket{\phi}_X\ket{\phi}_Y,
\end{equation}
where
\begin{equation}
\ket{\phi} = \sqrt{p}\,|0\rangle + \sqrt{1-p}\,|1\rangle,
\end{equation}
and measuring at the
intersection the observable
\begin{equation}
\sigma_z=|0\rangle\langle0|-|1\rangle\langle1|.
\label{sigmaz}
\end{equation}
When the driver arrives to any intersection, he measures
$\sigma_z$, and the probability of obtaining {\sc continue}
is~$p$, and the probability of obtaining {\sc exit} is~$1-p$.


\section{Entanglement-assisted strategies}


At this point, one can think about the possibility of using more
complex quantum mechanical resources instead of just experiments
on quantum systems as generators of random numbers. Indeed, here
we show that there are quantum strategies which give higher
expected pay-offs than any probabilistic one. For instance, let us
prepare two qubits in the singlet state
\begin{equation}
|\Psi\rangle = {1 \over \sqrt{2}} \left(|01\rangle-|10\rangle
\right), \label{singlet}
\end{equation}
and put one qubit in each intersection. The reduced states of each
qubit are both maximally mixed and cannot be described in terms of
classical correlations~\cite{Bell64}. When the driver arrives at
the first intersection, he measures $\sigma_z$, and the
probability of obtaining either {\sc continue} or {\sc exit}
is~1/2. However, by using the singlet state, the driver has
managed to introduce correlations between both intersections
without violating the assumptions of the absent-minded driver's
scenario: whenever he obtains {\sc continue} in the first
intersection, he obtains {\sc exit} in the second, and vice-versa.
Therefore, the expected pay-off is~2, which is higher than the
highest expected pay-off (4/3) for a probabilistic strategy. Note,
that as mentioned earlier, the use of pseudo-random (i.e.,
deterministic) procedures to generate anti-correlations would
immediately lead to distinguishable intersections: in principle
the driver could be in possession of the ``seed'' and would be
able to recognize the intersections by the {\em different}
outcomes of their pseudo-random number generators.

Moreover, the previous strategy achieves not only a higher pay-off
than any probabilistic strategy, but actually the optimal pay-off for the
absent-minded driver's problem. This can be proven as follows. The
driver has complete knowledge of the state of the universe before
the game starts, hence we can assume that the initial state is
pure $\ket{\Phi}_{XYR}$, where $R$ refers to the Hilbert space of
the rest of the universe. During the game he will have restricted
access to that state via sequential local operations on systems
$X$ and $Y$. Moreover, the driver lacks any memory or reference
that tells him which intersection he is at and therefore cannot
choose a strategy accordingly. With all the above considerations
we can now assume without loss of generality that an optimal
strategy for the absent-minded driver will consist on preparing
the quantum systems in $XY$ in a pure state $\ket{\Psi}$ and, on
his arrival at an intersection, performing a measurement in a
fixed basis, let us say the computational basis
$\{\ket{0},\ket{1}\}$. Indeed, such a scheme will always be able
to provide the same correlations between the measurement outcomes
than the most general scheme, and therefore provide the optimal
pay-off. We now have to maximize over the possible states
$\ket{\Psi}=\sum_{i,j=0}^{1}\alpha_{ij}\ket{ij}$. Since the
coherences do not affect the pay-off, it suffices to look at the
probability distribution $\{|\alpha_{ij}|^{2}\}$ of distinct
classical instruction sets $\{(i,j)\}$ and check for full
compliance of the rules when the phases are picked. Any
instruction that does not allow the driver to distinguish between
the intersections can be expressed as a convex linear combination
of three basic strategies. The first is the instruction to {\sc
continue} in both intersections (pay-off~$P_1=1$), represented by
$(0,0)$; the second is to {\sc exit} in both intersections
(pay-off~$P_2=0$), represented by $(1,1)$; and the third gives
{\sc continue} in $X$ and {\sc exit} in $Y$ for half of the runs,
and {\sc exit} in $X$ and {\sc continue} in $Y$ for the other half
(pay-off~$P_3=2$), and can be represented by $(0,1)+(1,0)$. The
optimal strategy should maximize $P=\sum_{j=1}^3 p_j P_j$, where
the probabilities are positive, $p_j \ge 0$, and normalized,
$\sum_{j=1}^{3} p_i =1$. Since the expected pay-off itself is a
convex function, it is clear that its maximum value will be
achieved with the extreme strategy of highest pay-off; that is,
the optimal pay-off is $P=P_3=2$ and can be reached with the
quantum state of Eq.~(\ref{singlet}).

A natural question is whether or not the same maximum pay-off
could be achieved with less quantum resources. Can we mimic the
strategy using the state~(\ref{singlet}) by means of a single
qubit (as a source of genuine randomness) coupled to a classical
mechanism to generate the desired correlations? The answer is no.
On one hand, obtaining the maximum pay-off does not require
perfect correlations but perfect {\em anti-correlations} between
the results on both intersections, and any classical mechanism
capable of generating different results would introduce a
distinction between the intersections. On the other hand, such a
classical mechanism {\em could} be used to transmit information
between the intersections when the driver is inside one of them
and thus violates (iii). Therefore we conclude that the strategy
using two qubits in the state~(\ref{singlet}) is the simplest
giving the maximum pay-off. At this point we also note that even
if one could justify the use of pseudo-random number generators
(or any other classical source of randomness), justifying the use
of two generators that create correlated random outputs is much
more demanding, as it immediately implies the existence of a
reproducible and hence predictable mechanism of generating the
outputs.


\section{The absent-minded driver's problem with
more intersections}


\begin{figure}
\centerline{\includegraphics[width=4.6cm]{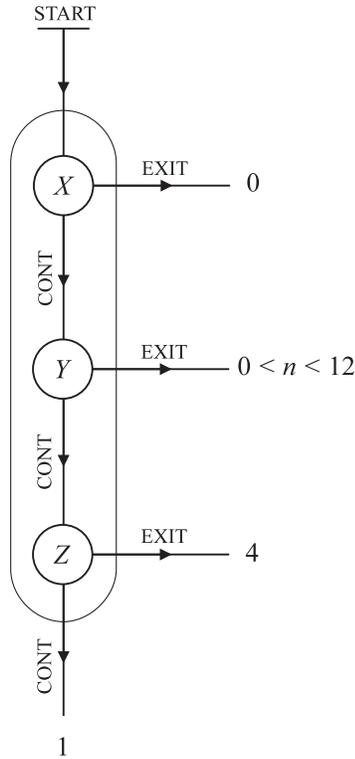}}
\caption{\label{Newroad} The absent-minded driver's problem with
three intersections.}
\end{figure}


\subsection{An example with three intersections}


Let us now consider the absent-minded driver's problem with $N >
2$ intersections. Here we can restrict ourselves to strategies
where, at each intersection, the driver measures in the
computational basis the prepared $N$-party pure state
$\ket{\Psi}=\sum_{\vec{v}}\alpha_{\vec{v}} \ket{\vec{v}}$, where
$\{\vec{v_{i}}\}_{1}^{d}$ is the set of $d=2^{N}$ $N$-dimensional
binary arrays, and corresponds to the set of instructions
\mbox{$\{(a,b,\stackrel{N}{\ldots}, e)\}$}. Here, finding the
optimal strategy also requires finding the vertices of the convex
polytope that results from imposing indistinguishability
conditions on the simplex probability space
$\{p_{1}=|\alpha_{v_{1}}|^{2},\ldots,
p_{d}=|\alpha_{v_{d}}|^{2}\}$. The condition for the
indistinguishability of the intersections is that the reduced
density matrices in each intersection have to be identical
$\rho_{1}=\ldots=\rho_{N}$. In the computational basis, the
diagonal matrix elements are equal if and only if all components
of the vector $\vec{u}=\sum_{m}p_{m}\vec{v}_{m}$ are identical,
i.e.,
\begin{equation}
\vec{u}(i)=\vec{u}(j) \; \; \forall
i,j=1,\ldots,2^{N}\mbox{.}
\label{eq:incond}
\end{equation}
For the moment we will disregard the conditions that arise from
equating the out-of-diagonal terms of the reduced density
matrices.

Any legitimate solution can be expressed as a convex combination
of the solutions associated to each vertex of the polytope. Each
vertex $E$ is specified by a vector $\vec{p}^{E}$ in the
probability space. As before, the optimal pay-off can be reached
at one of the vertices, and the corresponding mixed set of
instructions can be mapped into a quantum superposition by fixing
the arbitrary phases. The extremal property of the vertices
implies that all arrays contributing to a specific vertex, i.e.,
$\{\vec{v}_{m}: p_{m}^{E}\neq 0\}$ have to be linearly
independent. Using this fact and the indistinguishability
condition, it can be easily seen that two arrays contributing to a
vertex cannot differ solely in a single component. That is, for
any $(N-1)$-dimensional binary array $\vec{x}$, $(1,\vec{x})$ and
$(0,\vec{x})$ cannot appear together in the convex combination
that defines a vertex. From this it follows that all the
out-of-diagonal matrix elements of the reduced density operator
$\rho_{1}=\tr_{2,\ldots,N} \rho^{E}$ have to vanish,
\begin{equation}
\bra{1}\rho_{1}\ket{0}=\sum_{\{\vec{x}\}} \bra{1
\vec{x}}\rho^E\ket{0 \vec{x}}=0,
\end{equation}
where $\rho^{E}=\ketbrad{\phi^{E}}$ with
$\ket{\phi^{E}}=\sum_{m}\sqrt{p^{E}_{m}}\mathrm{e}^{i
\psi_{m}}\ket{\vec{v}_m}$. Therefore, out-of-diagonal terms will
not impose any additional constraints, and the optimal solution to
the absent-minded driver's problem can always be chosen to be
$\ket{\phi^{E}}=\sum_{m}\sqrt{p^{E}_{m}} \mathrm{e}^{i
\psi_{m}}\ket{\vec{v}_m}$ for a given vertex $\vec{p}^{E}$. A
remarkable feature is that all the vertices except two,
$(0,\ldots,0)$ and $(1,\ldots,1)$\textemdash which represent the
two possible deterministic strategies\textemdash, are implemented
by means of entangled states.

The particular solution for a given pay-off assignment can be
obtained in a straightforward manner by standard linear
programming methods. Obtaining the set of possible optimal
solutions can be significantly more difficult. For this purpose
one can use a method developed in the context of the theory of
$N$-party games. To every binary array $\vec{v}$, one can
associate a subset of $P=\{1,\ldots,N\}$,
$S_{v}=\{i:\vec{v}(i)=1\}$. The condition of indistinguishability
on the arrays $\{\vec{v}\}$ is equivalent, in game-theoretical
terms, to saying that the {\em collection} of sets, or coalition
of players, $\{S_{v}\}$ is {\em balanced}. The concept of a {\em
minimally balanced collection}~\cite{Scarf67,Shapley67}, which corresponds
to the set of contributing arrays in a vertex of the polytope, has
found important applications for the theory of $N$-party games.
This theory does not apply to the absent-minded driver's problem;
however, we can use an inductive method developed in that
context~\cite{Peleg65} to generate the possible solutions
(vertices) of the $N$-intersection problem from those of the $N-1$
problem.

\begin{figure}
\centerline{\includegraphics[width=12cm]{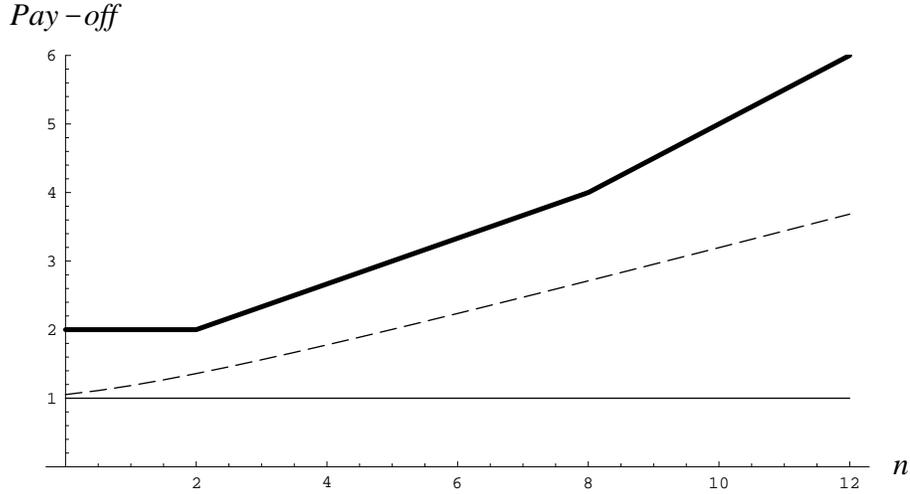}}
\caption{\label{Last} Optimal expected pay-offs as a function of
$n$ for deterministic (fine line, bottom), probabilistic (dashed
line, in the middle), and quantum entanglement-assisted (thick
line, top) strategies for the absent-minded driver's problem with
three intersections of Fig.~\ref{Newroad}.}
\end{figure}

Let us consider the absent-minded driver's problem
with three intersections~\cite{AHP97} and the set of pay-offs
given in Fig.~\ref{Newroad}. In this case, the vertices of the
polytope of possible solutions are $B_1=(0,0,0)$, $B_2=(1,1,1)$,
$B_3=(0,0,1)+(1,1,0)$, $B_4=(0,1,0)+(1,0,1)$,
$B_5=(1,0,0)+(0,1,1)$, $B_6=(0,0,1)+(0,1,0)+(1,0,0)$, and
$B_7=(0,1,1)+(1,0,1)+(1,1,0)$. It can be easily seen that, for any
$0 < n$, there is a probabilistic strategy that gives a higher
expected pay-off than any deterministic one (see Fig.~\ref{Last}).
More interestingly, there is a quantum entanglements-assisted
strategy that gives a higher pay-off than any probabilistic or
deterministic strategy (Fig.~\ref{Last}). For $0 < n \le 2$, the
optimum strategy (pay-off~$2$) can be implemented by preparing
three qubits in the so-called
Greenberger-Horne-Zeilinger~\cite{GHZ89} state
\begin{equation}
|{\rm GHZ}\rangle = {1 \over \sqrt{2}}
\left(|001\rangle+|110\rangle \right), \label{GHZ}
\end{equation}
putting one qubit in each intersection and measuring the
observable $\sigma_z$ defined in Eq.~(\ref{sigmaz}). For $2 \le n
\le 8$, the optimum strategy (pay-off~$(4+n)/3$) can be
implemented by preparing three qubits in the so-called
W~\cite{Pitowsky91,DVC00} state
\begin{equation}
|{\rm W}\rangle = {1 \over \sqrt{3}}
\left(|001\rangle+|010\rangle+|100\rangle \right), \label{W}
\end{equation}
followed by the $\sigma_z$ measurement. Finally, for $n \ge 8$,
the optimum pay-off (pay-off~$n/2$) can be reached in a similar
manner with GHZ-type states, for instance,
\begin{equation}
|{\rm GHZ'}\rangle = {1 \over \sqrt{2}}
\left(|011\rangle+|100\rangle \right).
\label{GHZ2}
\end{equation}

At the moment we do not have a characterization of the full set of
optimal solutions for the $N$-intersection problem. From the
previous reasoning we know that a necessary and sufficient
condition for a solution to be optimal is that the contributing
arrays are linearly independent. This also means that the
superpositions will involve at most $N+1$ terms. We also know
that, given a vertex (i.e., an optimal solution), if we exchange
$0$'s for $1$'s and vice versa, then we have another vertex. From
the previous examples it might seem that optimal solutions always
involve equally weighted superpositions. However, for $N>3$ one
can find optimal solutions for which this is not the case.


\section{Conclusions}


To sum up, the absent-minded driver's problem is that he does not
know which intersection he is at and cannot have instructions
indicating the route home. However, the assumptions under this
scenario do not keep him from using certain quantum superpositions
of instructions. We have seen that the use of entangled states
allows him to increase the expected pay-off beyond what was
previously considered the maximum pay-off.

The key point to understand the quantum advantage is that the
limited absent-mindedness of the driver requires even-in-principle
indistinguishability of the intersections; otherwise one cannot
exclude the possibility that the driver possesses some information
of the state of the rest of the universe (i.e., by pre-arranging
it before his trip) such that during his trip he can figure out
what intersection he is at. Hence, classically correlated physical
systems (and specially anti-correlated systems) that could provide
optimal expected pay-offs, are not allowed for decision-making.
However, quantum states such as~(\ref{singlet}), (\ref{GHZ}),
(\ref{W}), (\ref{GHZ2}), or those needed to implement all
non-trivial vertices guarantee indistinguishability {\em by
fundamental laws of physics}, while providing at the same time the
desired correlations. According to quantum mechanics the quantum
state of the $N$-qubit system is the most complete description of
the system, in the sense that it provides the maximum possible
information about any future experiments on these qubits. The
proposed entangled states, hence guarantee that the absent-minded
driver cannot determine at what intersection he is during his
trip. In addition, these states hide the ``correct'' instruction
sets in superpositions with non-optimal instructions, and thus
provide high pay-off values, without transgressing the assumptions
behind the absent-minded driver's scenario.

The above examples suggest that strategies based on quantum
entanglement can also provide advantages over classical strategies
in other decision problems involving memory limitations or
imperfect recall~\cite{PR97a,AHP97,PR97b} in which the
indistinguishability of the alternatives plays an essential role.


\section*{Acknowledgments}
We thank R.~Raussendorf for useful discussions, O.~G\"{u}hne for
reading the manuscript, and the organizers of the Ninth Benasque
Center for Science for support. A.C. acknowledges support from the
Spanish Mi\-nis\-te\-rio de Cien\-cia y Tec\-no\-lo\-g\'{\i}a
Grant No.~BFM2002-02815 and the Jun\-ta de An\-da\-lu\-c\'{\i}a
Grant No.~FQM-239.



\begin{thebibliography}{00}


\bibitem{PR97a}
M. Piccione, A. Rubinstein,
Games and Economic Behavior 20 (1997) 3.

\bibitem{AHP97}
R.J. Aumann, S. Hart, M. Perry,
Games and Economic Behavior 20 (1997) 102.

\bibitem{PR97b}
M. Piccione, A. Rubinstein,
Games and Economic Behavior 20 (1997) 121.


\bibitem{Klarreich01}
E. Klarreich,
Nature 414 (2001) 244.

\bibitem{VP02}
S.J. van Enk, R. Pike,
Phys. Rev. A 66 (2002) 024306.

\bibitem{Meyer99}
D.A. Meyer,
Phys. Rev. Lett. 82 (1999) 1052;
Phys. Rev. Lett. 84 (2000) 790.

\bibitem{EWL99}
J. Eisert, M. Wilkens, M. Lewenstein,
Phys. Rev. Lett. 83 (1999) 3077;
Phys. Rev. Lett. 87 (2001) 069802.

\bibitem{BH01}
S.C. Benjamin, P.M. Hayden,
Phys. Rev. Lett. 87 (2001) 069801;
Phys. Rev. A 64 (2001) 030301.

\bibitem{KJB01}
R. Kay, N.F. Johnson, S.C. Benjamin,
J. Phys. A 34 (2001) L547.

\bibitem{DLXSWZH02}
J. Du, H. Li, X. Xu,
M. Shi, J. Wu, X. Zhou, R. Han,
Phys. Rev. Lett. 88 (2002) 137902.

\bibitem{LJ02}
C.F. Lee, N.F. Johnson,
Phys. World 15 (2002) 25;
Phys. Rev. A 67 (2003) 022311.

\bibitem{GM03}
F. Guinea, M.A. Mart\'{\i}n-Delgado,
J. Phys. A 36 (2003) L197.


\bibitem{Bell64}
J.S. Bell,
Physics 1 (1964) 195.

\bibitem{GHZ89}
D.M. Greenberger, M.A. Horne, A. Zeilinger,
in: M. Kafatos (Ed.),
Bell's Theorem, Quantum Theory, and Conceptions of the Universe,
Kluwer, Dordrecht, 1989, p.~69.

\bibitem{Wheeler78}
J.A. Wheeler,
in: A.R. Marlow (Ed.),
Mathematical Foundations of Quantum Theory,
Academic Press, New York, 1978, p.~9.


\bibitem{Scarf67}
H.E. Scarf,
Econometrica 35 (1967) 50.

\bibitem{Shapley67}
L.S. Shapley,
Nav. Res. Logist. Q. 14 (1967) 453.

\bibitem{Peleg65}
B. Peleg,
Nav. Res. Logist. Q. 12 (1965) 155.


\bibitem{Pitowsky91}
I. Pitowsky,
Phys. Lett.~A 156 (1991) 137.

\bibitem{DVC00}
W. D\"{u}r, G. Vidal, J.I. Cirac,
Phys. Rev.~A 62 (2000) 062314.


\end{thebibliography}
\end{document}